\documentstyle[psfig]{ioplppt}
\jl{3}
\def\ke{\langle ke \rangle}
\def\pe{\langle pe \rangle}
\def\kvec{{\mathbf{k}}}
\def\mathbf#1{{\bf #1}}
\def\Im{\mathbf{Im}}
\def\Re{\mathbf{Re}}
\def\wp{\omega_{pl}}
\def\w{\omega}
\def\n0{n_0}
\def\v{\vec{v}}
\def\grad{\vec{\nabla}}
\def\dst{\displaystyle}
\def\etal{{\em et al}}
\def\vec#1{{\bf #1}}

\begin{document}
\letter{Exchange--correlation potential for Current Density Functional
Theory of frequency dependent linear response} 
\author{S Conti, R Nifos\`i and M P Tosi}
\address{INFM and Classe di Scienze, Scuola Normale Superiore, 
Piazza dei Cavalieri, 
I-56126 Pisa, Italy}

\begin{abstract}
The dynamical, long--wavelength longitudinal and transverse
exchange--correlation potentials for a homogeneous electron gas are
evaluated in a microscopic model based on an
approximate decoupling of the equation of motion for the
current--current response function. The transverse spectrum turns out
to be very similar to the longitudinal one.
We obtain evidence for a strong spectral structure near twice the
plasma  frequency  due to a two--plasmon threshold
for two--pair excitations, which may be observable in inelastic
scattering experiments.
Our results give the entire input needed to implement the 
Time--Dependent Current Density Functional Theory scheme 
recently developed by G. Vignale and W. Kohn [Phys. Rev. Lett. {\bf 77},
2037  (1996)] and are fitted to analytic functions to facilitate such
applications.
\end{abstract}


The Time--Dependent Density Functional Theory (TD--DFT) of Runge and
Gross \cite{RungeGross} in principle allows the study of dynamical
properties of interacting many--particle systems, which are not
addressed by static DFT. 
The adiabatic local density approximation (ALDA) \cite{Zangwill80}
has enabled successful application  of the theory in the
low--frequency limit.
The search for a local dynamical approximation to the
exchange--correlation (xc) potential
has been frustrated  by the appearence of various inconsistencies
\cite{Dobson94,Vignale95b,Gross96}, which can be tracked down
to the non--existence of a gradient expansion for the
frequency--dependent xc potential in terms of the density alone. 

Recently Vignale and Kohn (VK) have shown how a local approximation for the
xc vector potential ${\bf a}^{xc}(\vec{r},\w)$ can
be obtained within linearized Current Density Functional
Theory \cite{Vignale96,VignaleKohn96b}. They prove that for slowly
varying densities 
known symmetries and conservation laws allow one to express
${\bf a}^{xc}(\vec{r},\w)$ exactly in terms 
of the long--wavelength xc potentials
$f_{xc}^{L,T}(\w)$ for a homogeneous electron gas,
\begin{eqnarray}
\fl
{\bf a}^{xc}_{i}(\vec{r},\w)=  
-\frac{1}{\w^{2}}\left\{\partial_{i}[f^{L}_{xc}\grad\cdot(n_0\v)-
\delta f^{L}_{xc}\v\cdot\grad n_0]+\delta f^{L}_{xc}(\partial_i\n0)
\grad\cdot\v\nonumber
\right. \\
 + f^{T}_{xc}\left[-\n0(\vec{\nabla}\times\vec{\nabla}\times\v)_{i}
+2(\partial_j \vec{v}_i + \partial_i \vec{v}_j) \partial_j n_0
- 4 (\partial_i n_0) \nabla\cdot\vec{v}\right]\nonumber  \\
 + \left.\n0\left(
(\partial_j f^{T}_{xc})\left(\partial_i\v_j+\partial_j\v_i\right)
-2(\partial_i f^{T}_{xc})\nabla\cdot\vec{v}\right)\right\}\;.
\label{axc}
\end{eqnarray}
Here, $\n0(\vec{r})$ is the unperturbed ground--state density,
$\v(\vec{r},\w)=\vec{j}_1(\vec{r},\w) /\n0(\vec{r})$ is the local
velocity and $\delta f_{xc}^L(\w,n) = f_{xc}^L(\w,n) - f_{xc}^L(0,n)$.
The potentials $f_{xc}^{L,T}$, which are functions of $\omega$  
and of the local density $n_{0}(\vec{r})$, are defined as
the $k\to0$ limit of
\begin{equation}
\label{eqdeffxltchi}
f_{xc}^{L,T}(k,\w) = {\w^2\over k^2} \left\{
 {1\over \chi^0_{L,T}(k,\w)+\rho/m}-
{1\over\chi_{L,T}(k,\w)+\rho/m} \right\} -v_{L,T}
\end{equation}
where $\chi_{L(T)}$ is the current--current longitudinal
(transverse) response function of
the homogeneous system at density $\rho=n_0(\vec{r})$, $\chi^0_{L(T)}$
is the equivalent ideal--gas quantity, $v_L = 4\pi e^2/k^2$ and $v_T=0$.

The quest for a frequency--dependent local field factor
$G(k,\w)=-f_{xc}^L(k,\w)/v_L$  for the density--density response is a  
long--lasting problem in electron gas theory
\cite{SingwiTosi,GrossKohn85,IGK,HW,Holas89,GlickLong,Richardson94,Neilson}.
Its asymptotic behaviours are known 
from sum rule arguments \cite{SingwiTosi,GrossKohn85}
and from second--order perturbative expansions
\cite{HW,Holas89,GlickLong}, and a smooth interpolation scheme was
proposed by Gross and Kohn  
\cite{GrossKohn85,IGK}.
A fully microscopic result, derived from an 
approximate treatment of two--pair excitations, 
was obtained by B\"ohm \etal \ \cite{BCT96,BCT96b} using an expression
first derived by Hasegawa and Watabe \cite{HW}. 
The longitudinal component of the current--current xc 
potential $f_{xc}^L(\w)$ can be obtained from such results, 
but no sensible approximations have 
been proposed for the transverse component, preventing implementation
of the Vignale--Kohn theory.

In this letter we develop a full treatment of the two--pair excitation
spectrum, including its transverse component. Our results 
give the entire input necessary to use the VK expression.

The xc potential
can be obtained within the equation of motion formalism. A general
response function is defined as
\begin{equation}
\ll A;B \gg_\w\,=-i\int_{0}^{\infty}dt e^{i(\omega+i\epsilon)t}<[A(t),B(0)]>
\end{equation}
where $A(t)=e^{iHt}Ae^{-iHt}$ and $<\dots>$ denotes ground state
expectation values.
The current--current response 
$\chi_{ij}(k,\omega)=\,\ll\vec{j}_{\kvec}^{i};\vec{j}_{-\kvec}^{j}\gg$
satisfies the equation of motion
\begin{equation}
\chi_{ij}(k,\w)= {1\over\w^2} <\left[[\mathbf{j}_{\bf k}^i,
H],\mathbf{j}_{\bf -k}^j\right]> 
-{1\over\omega^2} \ll [\mathbf{j}_{\bf k}^i,H];[\mathbf{j}_{\bf -k}^j,H] 
\gg_\w\,. 
\label{eqeqmotchiij}
\end{equation}
The last term yields four--point
response functions $\ll AB;CD \gg_\w$ involving $\rho_{\mathbf{k}}$ and
$\mathbf{j_{k}}$ operators and can
be approximately decoupled by an RPA--like scheme:
\begin{eqnarray}
\Im\ll AB;CD \gg_{\omega}&\simeq&-\int_0^\omega
 \frac{d\omega'}{\pi}[\Im\ll A;C  \gg_{\omega'}
\Im\ll B;D \gg_{\omega-\omega'}\nonumber\\
&& +\Im\ll  A;D \gg_{\omega'}
\Im\ll B;C \gg_{\omega-\omega'}].
\label{eqdecouplingabcd}
\end{eqnarray}
Such a decoupling includes by construction two--pair processes, which
are the lowest order processes with non--zero spectral strength in the
relevant region of the $(k,\w)$ plane. Perturbative
approaches evaluate the LHS of equation (\ref{eqdecouplingabcd}) for an
ideal gas and obtain a spectrum restricted to single pair
excitations, which is essentially equivalent to the one obtained by using
non--interacting response functions in the RHS (the difference lies in
exchange processes, as discussed below). We intend to include the
effect of plasmons and therefore use RPA response functions in the
RHS.

\begin{table}
\caption{Exact limiting behaviours of $f_{xc}^{L,T}(\w)$ from the
Monte Carlo equation of state, in units of $2\wp/\rho$, and best fit
parameters for $\Im\,f_{xc}^L$ according to (\protect\ref{eqfitfxc}).} 
\footnotesize\rm
\label{tablefxcasy}
\label{tabellafitimfxcL}
\begin{center}
\begin{tabular}{@{}lllllllllll}
\br
$r_s$ & $f_{xc}^L(0)$ &  $f_{xc}^L(\infty)$ & 
$f_{xc}^T(\infty)$ & $\beta$ & $10^2 c_0$ & $10^2 c_1$ &  $\w_1$ & $\w_2$ &
 $d_0$ & $10^2d_1$ \\ \mr
0.5  & -0.04246 & -0.01794 & 0.0177 & 1.87&  0.175 & 0.694 &  1.75 &
-3.59 & 0.173 & 5.72\\ 
1  & -0.0611 & -0.0216 & 0.0284  &  1.48 &  0.421 &  1.76 &  0.982 &
-1.45 & 0.291 & 9.38\\ 
2  & -0.0891 & -0.0252 & 0.0457  &  1.22 &  0.895 &  3.87 &  0.347 &
0.181 & 0.49 &  13.2\\ 
3  & -0.1119 & -0.0280 & 0.0600  &  1.1 &   1.29 &  6.09 &  0.143 &
0.693 &  0.664 & 16.7\\ 
4  & -0.1320 & -0.0308 & 0.0724  &  1.02 &  1.65 &  7.87 &  -0.143 &
1.33 &  0.824 & 17\\ 
5  & -0.1503 & -0.0338 & 0.0835  &  0.955 & 1.94 &  9.82 &  -0.27 &
1.61 &  0.974 &  18.3\\ 
6  & -0.1674 & -0.0370 & 0.0935  &  0.899 & 2.22 &  11.6 &  -0.361 &
1.82 & 1.12 &  19.3\\ 
10  & -0.2276 & -0.0518 & 0.1267  &  0.698 &3.11 &  17.9 &  -0.565 &
2.27 & 1.64 &  22.1\\  
15  & -0.2917 & -0.0725 & 0.1587  &  0.474 &3.94 &  24.2 &  -0.69 &
2.54 &  2.22 &  23.9\\ 
20  & -0.3483 & -0.0939 & 0.1847  &  0.259 &5.54 &  24.7 &  -0.808 &
2.78 & 2.75 &  22.8\\  
\br
\end{tabular}
\end{center}
\end{table}

Quite lengthy calculations lead in this approach to
the following result for the xc
potentials $f^{L,T}_{xc}(\w)$ in terms of the response functions
introduced in equation 
(\ref{eqdeffxltchi}): 
\begin{eqnarray}
{\mathbf{Im}}f_{xc}^{L,T}(\omega) &=&
-
\int^\omega_0{d\omega'\over\pi}
\int{d^3q\over(2\pi)^3\rho^2} v^2_{q}
{q^2\over(\omega-\omega')^2} {\mathbf{Im}}\chi_L(q,\omega-\omega')
\nonumber\\
&&\times
\left[a_{L,T}{q^2\over\omega'^2}{\mathbf{Im}}\chi_L(q,\omega')+
b_{L,T}{q^2\over\omega^2} \mathbf{Im}\chi_T(q,\omega')\right]
\label{eqimchil}
\end{eqnarray}
with $a_L=23/30$, $a_T=8/15$, $b_L=8/15$ and $b_T=2/5$. 
The expression for the longitudinal part is equivalent to the
one obtained by Hasegawa and Watabe \cite{HW} by diagrammatic means, and 
similar to the one obtained by Neilson {\em et al.} \cite{Neilson}
(see the discussion in \cite{BCT96}). The
result for the transverse component is new.

The real part of $f_{xc}$ can be obtained via the Kramers--Kronig
relations, namely 
\begin{equation}
{\mathbf{Re}} f_{xc}(\omega)={\mathbf{Re}}f_{xc}(\infty)+
\frac{1}{\pi} \int_{-\infty}^{+\infty}d\omega'
\frac{{\mathbf{Im}}f_{xc}(\omega')}
{\omega'-\omega}\,.
\label{KK1}
\end{equation}
The first term in the RHS of equation (\ref{eqeqmotchiij}) fixes the
high--frequency limit
\begin{eqnarray}
f_{xc}^{L,T}(\omega=\infty) = 
\frac{1}{2\rho}\left[d_{L,T}(\ke-\ke^0)+e_{L,T}\pe\right]
\label{eqfxclwinfty}
\end{eqnarray}
with $d_L=4$, $d_T=4/3$, $e_L=8/15$ and $e_T=-4/15$. 
The average kinetic and potential energies $\ke$ and $\pe$ can be
obtained from the Monte Carlo equation of state \cite{CeperleyAlder},
as can the compressibility $K_T$ which fixes the longitudinal static limit
\begin{equation}
\lim_{k\to0}\lim_{\omega\to0}f^{L}_{xc}(k,\omega) =
\frac{1}{\rho^2} \left(\frac{1}{K_T} -
\frac{1}{K^0_T}\right)\,.
\label{eqfxclw0}
\end{equation}
In the above equations, $\ke^0$ and $K_T^0$ denote ideal gas results.
The asymptotic values of $f_{xc}^{L,T}$ are listed in Table
\ref{tablefxcasy}. 
We remark that there is no proof that the order of limits in
equation (\ref{eqfxclw0}) can be safely interchanged nor in general that
$f_{xc}^L(k,\w)$ is a continuous function in the limit
$(k,\w)\to(0,0)$. Indeed, there 
are arguments which indicate a small 
discontinuity \cite{footnotefxc}.  
In view of (i) the small estimated discontinuity, (ii) the 
uncertainty in its precise value and (iii) the
appeal of a theory which reduces continuously to LDA as $\w\to0$, we
prefer to enforce continuity as explained below.  

\begin{figure}[p]
\centerline{
\psfig{figure=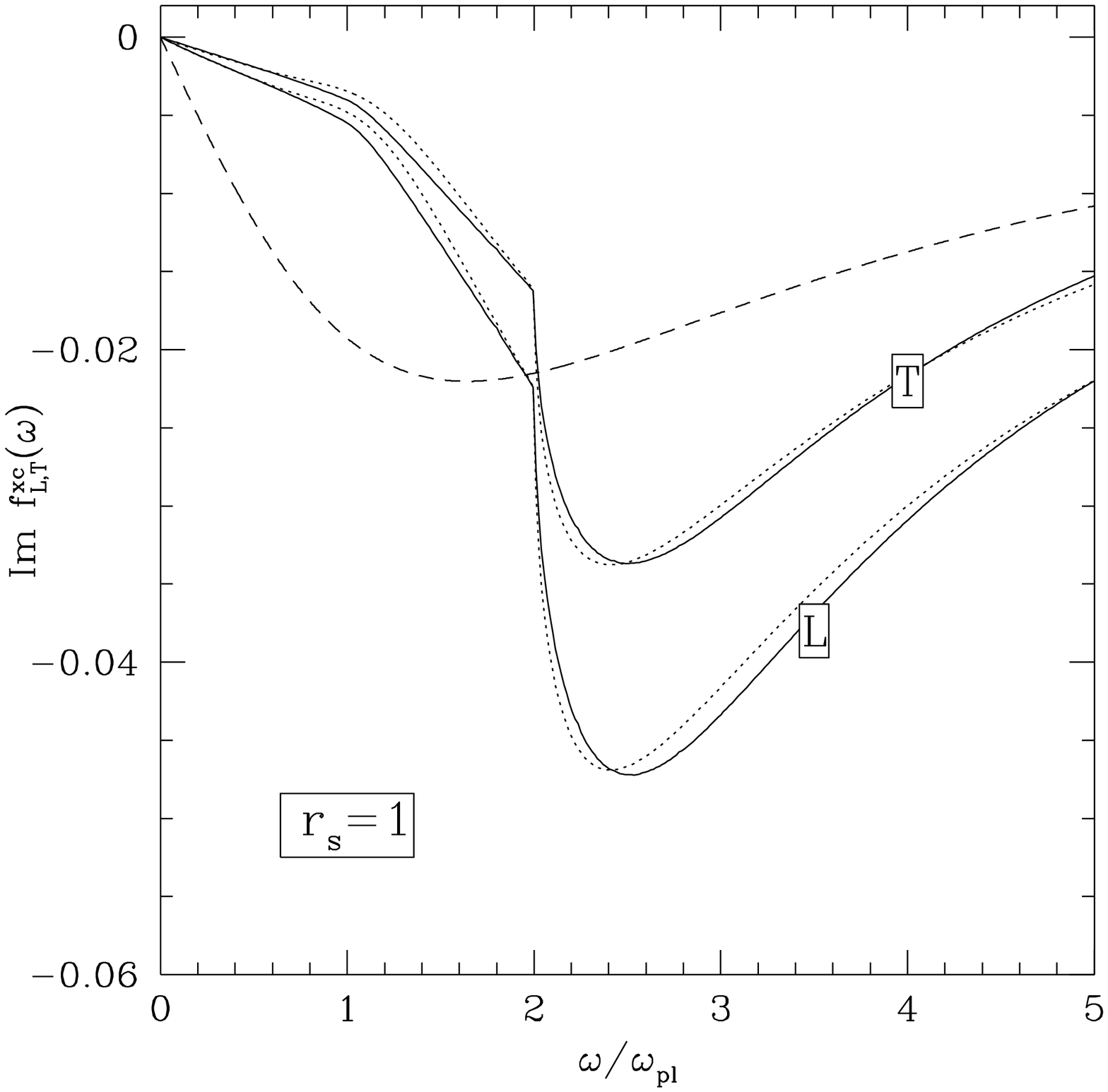,width=0.43\linewidth,rheight=0.4\linewidth}
\psfig{figure=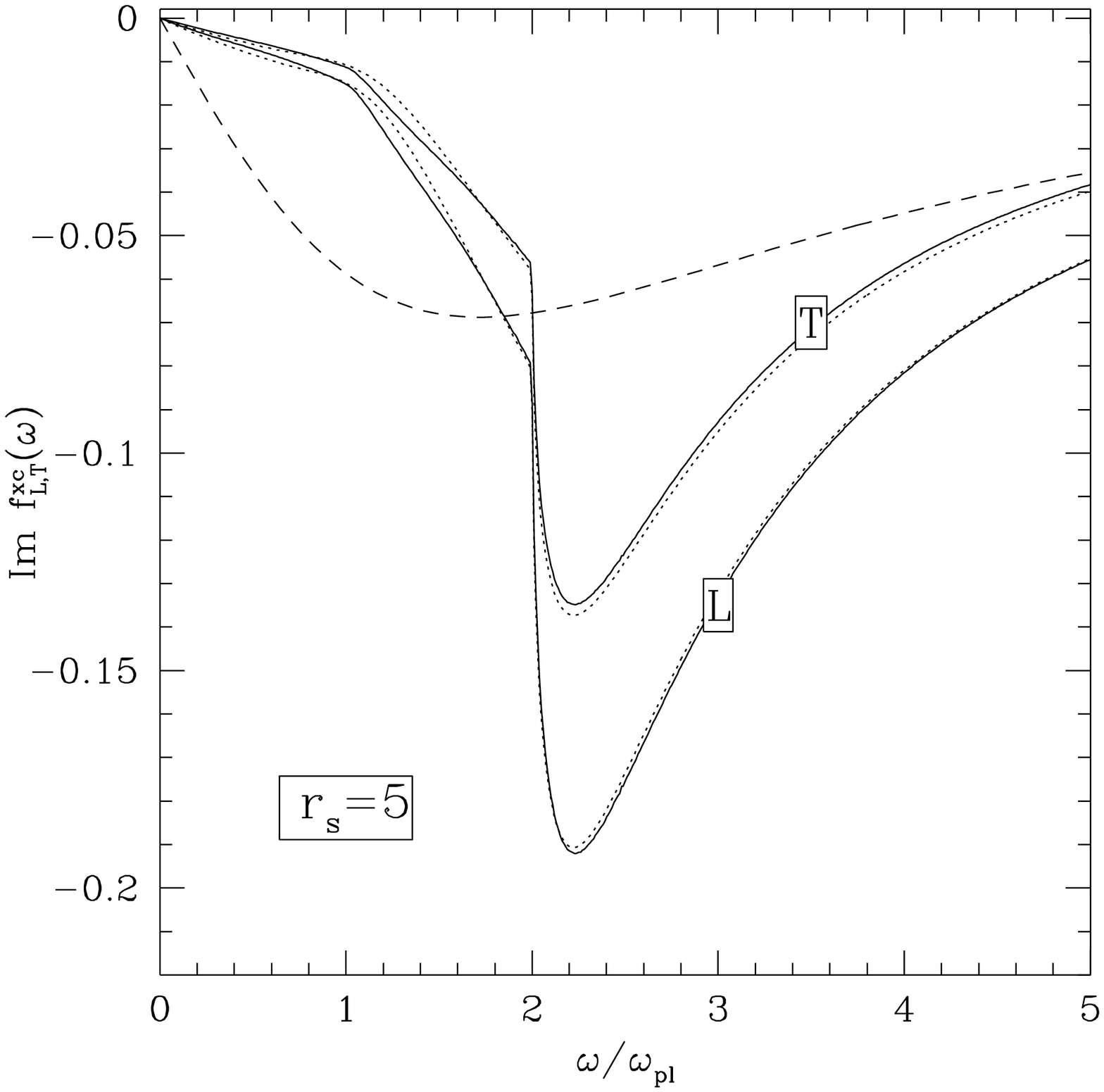,width=0.43\linewidth,rheight=0.4\linewidth}}
\caption{Imaginary parts of $f^L_{xc}(\w)$ and  $f^T_{xc}(\w)$ in
units of $2\wp/\rho$,  as functions of $\w/\wp$ at 
$r_s=1$ (left panel) and $r_s=5$ (right panel). The present results
(full curves) are 
compared with the Gross--Kohn interpolation for the longitudinal
component (dashed curves) and with the fit discussed in the text
(dotted curves).}
\label{figim}
\end{figure}

The above approximation neglects exchange processes,  which in
perturbative 
treatments reduce the total two--pair spectral weight by a factor of 2
if $\w\gg \varepsilon_F$, $\varepsilon_F$ being the Fermi energy. 
We approximately include exchange by multiplying
$\Im f_{xc}^L$ by the phenomenological factor
\begin{equation}
g_x(\w)={\beta  + 0.5 \w/2\varepsilon_F
\over 1+ \omega/2\varepsilon_F}\,,
\label{eqdefgx}
\end{equation}
where the parameter  $\beta$ is used to enforce continuity at $k=\w=0$
and turns out to be close to 1 at
metallic densities (see Table \ref{tabellafitimfxcL}).
For consistency we use the same 
factor $g_x(\w)$ in the transverse component. Our final
high--frequency result is 
\begin{equation}
\Im f_{xc}^{L,T} (\w\to\infty) = -2 c_{L,T} \left(2 Ry \over
\w\right)^{3/2} a_B^3 Ry 
\label{eqfxcltgllargew}
\end{equation}
where $c_L=23\pi/15$ and $c_T=16\pi/15$, in agreement with 
the result of Glick and Long \cite{GlickLong} 
for the longitudinal term and 
its extension to the transverse one (unpublished). 

Equations (\ref{eqimchil}) could be considered as the $k=0$
component of a self--consistency condition, but the self--consistent
problem cannot be solved since the equations determining the
finite--$k$ behaviour of $f_{xc}$ are unknown. 
We have evaluated equation  (\ref{eqimchil})  
by using the RPA response functions in the LHS. 
The imaginary part of the RPA longitudinal susceptibility consists of
two separate  contributions: the first arises from the broad single--pair 
continuum, the other from  the sharply peaked plasmon excitation. In
turn this leads to a structure in $\Im f_{xc}(\w)$ around twice the
plasma frequency (for details see Ref. \cite{BCT96}). As we are
neglecting retardation the transverse
susceptibility contains only a broad continuum and therefore has 
larger spectral strength than the longitudinal one at low frequency.

\begin{figure}[p]
\centerline{
\psfig{figure=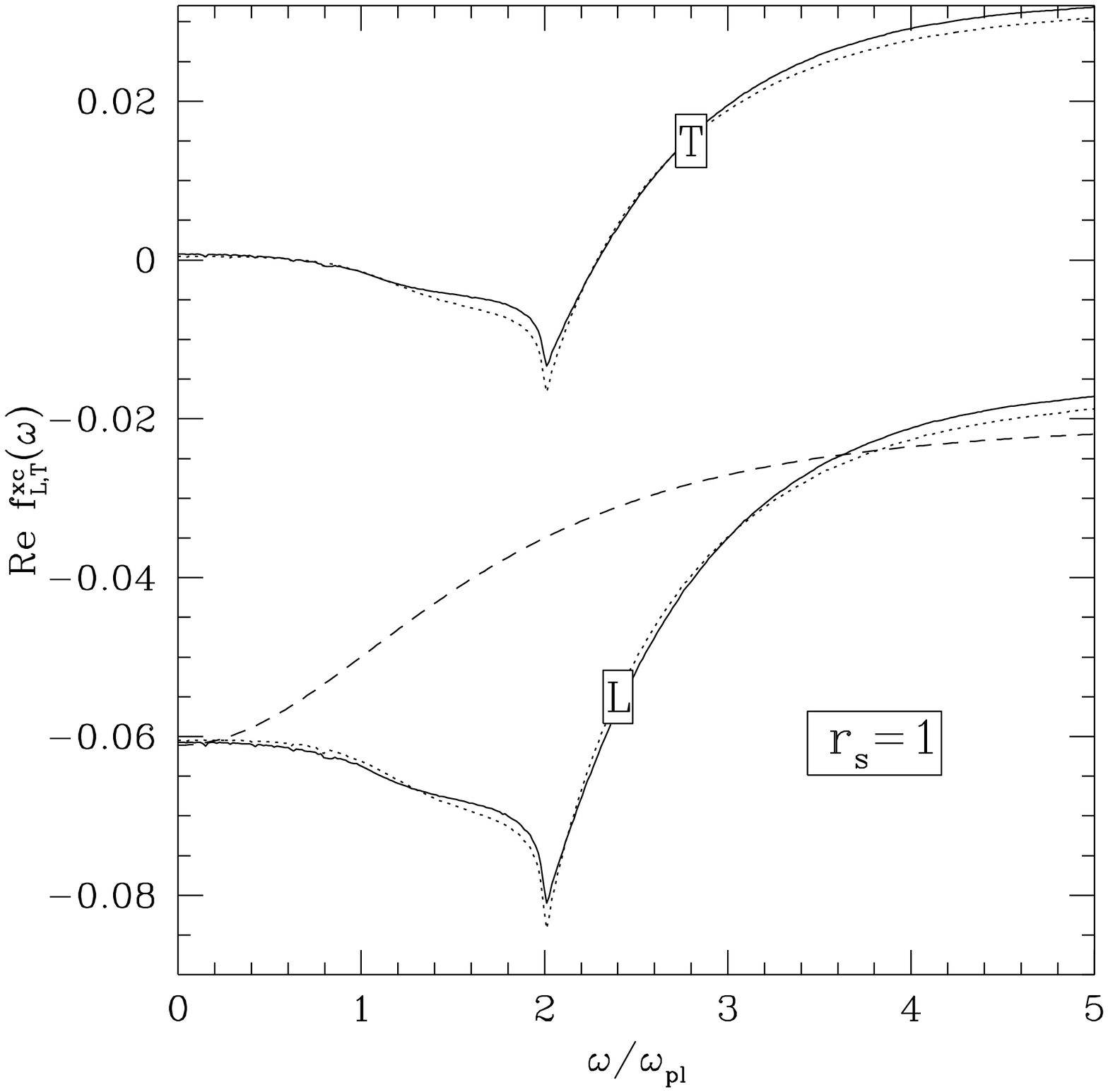,width=0.43\linewidth,rheight=0.4\linewidth}
\psfig{figure=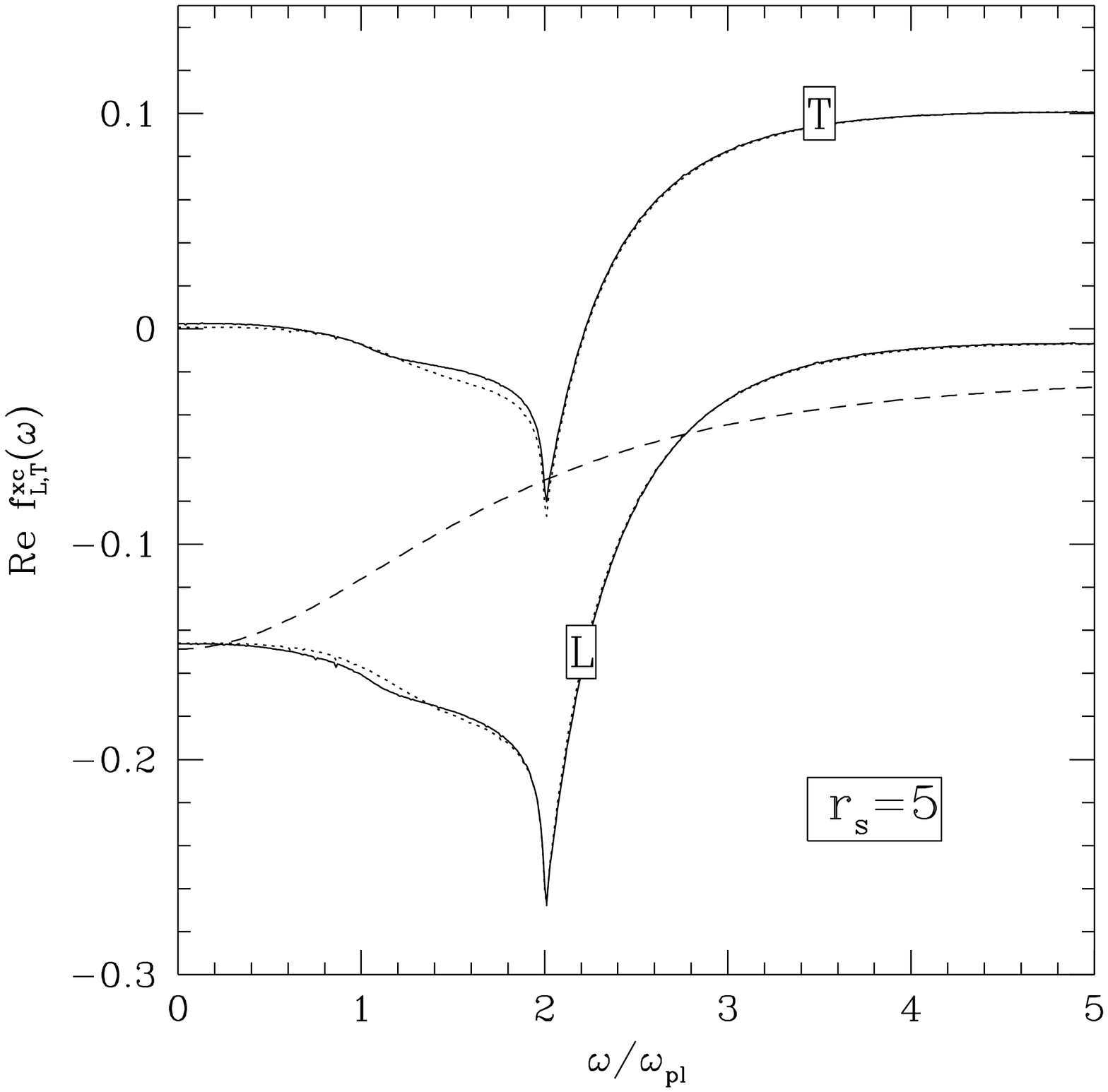,width=0.43\linewidth,rheight=0.4\linewidth}}
\caption{Real parts of $f^L_{xc}(\w)$ and  $f^T_{xc}(\w)$
(full curves), in units of $2\wp/\rho$, as functions of $\w/\wp$ at 
$r_s=1$ (left panel) and $r_s=5$ (right panel). The meaning of the
other curves is as in Figure \protect\ref{figim}.}
\label{figre}
\end{figure}

Figures \ref{figim} and \ref{figre} report our results for the
imaginary and real parts of $f_{xc}$  and compare them 
with the interpolation scheme of Gross and Kohn \cite{GrossKohn85} for
the longitudinal term. 
Both curves reproduce the asymptotic limits (\ref{eqfxclwinfty}) and
(\ref{eqfxclw0}) as well as the $\w^{-3/2}$ high--frequency behaviour,
but the behaviours at intermediate frequencies are strikingly
different. Our curves for $\Re f_{xc}$ exhibit a sharp
minimum around 
$2\wp$, which corresponds to the sharp structure found in $\Im f_{xc}$
at the same frequency. The physical origin lies in the large spectral
strength of the plasmon excitation as compared to single--pair
excitations, which accumulates most of the spectral strength of
two--pair processes near $2\wp$. This spectral structure becomes
sharper with increasing coupling strength, i.e. with increasing $r_s$
($r_s$ is defined as $(4\pi\rho a_B^3/3)^{-1/3}$, $a_B$ being the
effective Bohr radius). We note that within the accuracy of the
present model $f_{xc}^T(\w=0)$ is indistinguishable from zero in the
entire density range explored here.
 
Figure \ref{figre} shows stronger xc effects in the real part at
intermediate frequencies ($\w\simeq\wp$) than in both the $\w=0$ and the
$\w=\infty$ limit, at variance from the Gross--Kohn
interpolation. Available experimental data on the plasmon dispersion
coefficient in alkali metals \cite{Fink} indicate much stronger
deviations from RPA than predicted by existing electron
gas theories. Although band--structure effects play a significant
role in a quantitative comparison, the dynamic correction
indicated by our results improves the agreement with the
experimental data.

\begin{figure}[p]
\centerline{%
\psfig{figure=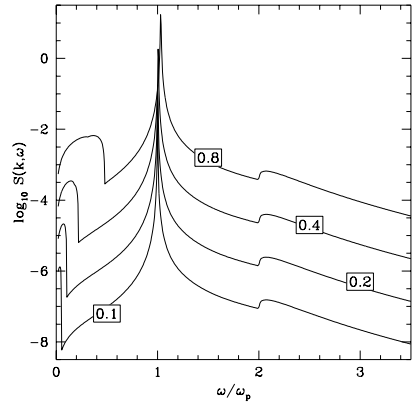,width=0.43\linewidth,rheight=0.4\linewidth}}
\caption{Dynamic structure factor $S(k,\w)$ at $r_s=5$ as a function
of $\w/\wp$ at various values of $k r_s a_B$ on a semilogarithmic
scale.}
\label{figskw}
\end{figure}

Figure \ref{figskw} displays the dynamic structure 
factor $S(k,\w) = -(2k^2/\rho\w^2) \Im \chi_L(k,\w)$
which is relevant to inelastic scattering experiments. The threshold
behaviour at frequency $2\wp$ is a clearcut signature
of the present results for $f_{xc}$.

While the detailed shapes of our results are largely
approximation--dependent, we do not expect the main qualitative features to
be substantially modified in a more refined theory. Both self--consistent SPA
results \cite{BCT96} and preliminary calculations with an STLS
response function \cite{SingwiTosi}  show the same qualitative
behaviours at values of $r_s$ where the plasmon
dispersion coefficient is positive. A much deeper
minimum in $\Re f_{xc}$ is found at negative plasmon dispersion,
which in STLS is the case for $r_s>5$. 

We also remark that the shapes of the longitudinal and transverse
spectra are very similar. Equation (\ref{eqimchil}) shows that $\Im
f_{xc}^T = 16/23\cdot \Im f_{xc}^L$ for $\w\gg\wp$ whereas $\Im
f_{xc}^T = 3/4\cdot \Im f_{xc}^L$ for $\w\ll\wp$. 
In fact, the
transverse spectrum is rather accurately reproduced at all frequencies
by setting $\Im f_{xc}^T(\w) \simeq 0.72 \cdot \Im f_{xc}^L(\w)$, whereas for
the real part there is an additional shift due to the different
$\w=\infty$ value.

Our numerical results for $\Im
f_{xc}^L(\w)$ can be accurately reproduced by the expression 
\begin{equation}
\Im f_{xc}^L(\w)= -g_x(\w) \left\{
\begin{array}{ll}
\dst c_0 \w + c_1{\w-1\over
e^{(7/\w)-5}+1} \hskip1cm & (\w<2) \\[5mm]
\dst {d_0 \sqrt{\w-2} + d_1\over \w (\w - \w_1\sqrt{\w} - \w_2)}
& (\w>2) \end{array} \right.
\label{eqfitfxc}
\end{equation}
where $\w$ is in units of $\wp$, $f_{xc}^L$ in units of
$2\wp/\rho$.
The fit parameters given in Table \ref{tabellafitimfxcL} were 
obtained by imposing (i) continuity at $\w=2\wp$, (ii) conservation of the
normalization $\int \Im f_{xc} d\w/\w$ and (iii) the asymptotic
behaviour given by equation (\ref{eqfxcltgllargew}). The
remaining three parameters were fitted to the numerical data. 
The real part can be obtained from the  Kramers--Kronig
relation and the high--frequency values given
in Table \ref{tablefxcasy}; the low--frequency longitudinal values
from the same Table can be used for a check. The good
quality of the  resulting fits is shown in Figures \ref{figim} and
\ref{figre}.   

In summary, we have presented a microscopic model for the longitudinal and
transverse xc potentials of the 3D homogeneous
electron gas. We have given numerical evaluations yielding results
which are significantly different from previous interpolation
schemes and compatible with experimental
data on the plasmon dispersion coefficient. Our results have been
fitted to analytic 
expressions and provide all the input needed for TD--DFT
computations in the linearized long--wavelength regime.

\ack
We gratefully acknowledge very useful discussions with
Professor G. Vignale.

\end{document}